# First-principles study of the lattice dynamical properties of strontium ruthenate

Naihua Miao,[1,*] Nicholas C Bristowe,[1] Bin Xu,[2] Matthieu Verstraete[2] and Philippe Ghosez[1]

[1]*Physique Théorique des Matériaux, Institut de Physique, Université de Liège, B-4000 Sart Tilman, Belgium*

[2]*Physique des Matériaux et Nanostructures, Institut de Physique, Université de Liège, B-4000 Sart Tilman, Belgium*

[*]Corresponding author: nhmiao@gmail.com or naihua.miao@ulg.ac.be

Abstract:

By means of first-principles calculations, various properties of $SrRuO_3$ are investigated, focusing on its lattice dynamical properties. Despite having a Goldschmidt tolerance factor very close to 1, the phonon dispersion curves of the high-temperature cubic phase of $SrRuO_3$ show strong antiferrodistortive instabilities. The energetics of metastable phases with different tilt patterns are discussed, concluding that the coupling of oxygen-rotation modes with anti-polar Sr motion plays a key role in stabilizing the *Pnma* phase with respect to alternative rotation patterns. Our systematic analysis confirms previous expectations and contributes to rationalize better why many $ABO_3$ perovskites, including metallic compounds, exhibit an orthorhombic ground state. The zone-center phonon modes of the *Pnma* phase have been computed, from which we propose partial reassignment of available experimental data. The full dispersion curves have also been obtained, constituting benchmark results for the interpretation of future measurements and providing access to thermodynamical properties.

1. Introduction

Among complex oxides, ABO$_3$ perovskites have received continuous attention over the last few decades since they can develop various types of ordering of their charge, spin and orbital degrees of freedom and exhibit a wide range of functional properties including ferroelectricity, piezoelectricity, (anti-)ferromagnetism, multiferroism, metal-insulating transitions, and superconductivity [1]. Thanks to the recent advances of experimental and theoretical techniques, the interest in this class of material has not been restricted to bulk compounds but also extended to nanostructures. During the last few years, it was discovered that exciting new phenomena can emerge at the interfaces in multilayer heterostructures and superlattices [2,3].

SrRuO$_3$, is a well-known metallic perovskite compound, widely used as an electrode material [4-6] and also appears as the simplest member of the interesting Ruddlesden-Popper series of layered ruthenates Sr$_{n+1}$Ru$_n$O$_{3n+1}$. It has fascinated researchers for many years owing to its surprising itinerant ferromagnetism and unusual transport properties. It has also generated a debate regarding its degree of electronic correlation and related consequences [7].

Like many perovskites, bulk SrRuO$_3$ undergoes a sequence of consecutive structural phase transitions with decreasing temperature [7-10]. The high symmetry cubic phase ($Pm\bar{3}m$) is stable only above 950 K [8-10]. At 950 K, the cubic phase first turns to a tetragonal structure (*I*4/*mcm*) which was observed experimentally within a very narrow range between 950 K and ~825 K [8-10]. Around 825 K, the tetragonal SrRuO$_3$ transforms to a *Pnma* ground state either directly [8,10] or, as suggested more recently, via a structure with *Imma* symmetry, though the intermediate *Imma* phase has only been observed between 825K and 685K by two experiments to date [9,11]. The orthorhombic *Pnma* phase of SrRuO$_3$ has a GdFeO$_3$-type structure [12] and is a metallic ferromagnet below a Curie temperature of ~160 K. In line with previous works by Thomas [13] and Woodwards [14], Benedek and Fennie recently [15] confirmed at the first-principles level that anti-polar A-cation motions significantly contribute to stabilize a *Pnma* ground-state in perovskites. In this latter work, they focused on a series of insulating ABO$_3$ compounds. It might be interesting to quantify how this explanation applies to metallic systems like SrRuO$_3$.

Various physical properties of $SrRuO_3$ have already been explored theoretically in the past. Early work by Singh [16], Allen [17], Santi [18] and Shepard [19] and their coworkers discussed the electronic, structural, transport, thermodynamic and magnetic properties of bulk $SrRuO_3$. The structural, electronic and magnetic properties of $SrRuO_3$ under epitaxial strain have also been studied theoretically and experimentally [20-23]. Toyota *et al* [24] and Rondinelli *et al* [25] investigated the thickness dependent metal-to-insulator transition in thin films. The possibility of creating a spin-polarized 2-dimensional electron gas highly-confined in the $SrRuO_3$ layers of $(SrTiO_3)_5/(SrRuO_3)_1$ superlattices has been predicted theoretically [6] and the related thermoelectric properties of the system have been discussed [26]. Although it was often assumed to be a strongly correlated system, it was recently suggested that $SrRuO_3$ should be considered as a weakly correlated itinerant magnet [27,28].

Compared to the efforts devoted to the understanding of its structural, electronic and magnetic properties, very little has been done regarding the characterization of the dynamical properties of $SrRuO_3$. Experimental Raman and infrared (IR) spectroscopy measurements for orthorhombic *Pnma* $SrRuO_3$ thin films have been performed, but these are not complete and many modes were not detected or assigned (as discussed in Reference [7]). Moreover, the dynamical properties of the high symmetry cubic $SrRuO_3$ have not been reported, which motivates us to explore the dynamical properties of $SrRuO_3$ systematically.

In the present work, by utilizing first-principles calculations, we study the structural, electronic, magnetic and dynamical properties of different phases of $SrRuO_3$. The full phonon dispersion curves of the cubic phase are reported, highlighting the presence of a branch of antiferrodistortive (AFD) phonon modes extending from the *R* to the *M* points of the Brillouin zone. We then investigate several possible intermediate phases arising from the condensation of different combinations of antiferrodistortive phonon modes. We compare their internal energies and shed light on the microscopic mechanism stabilizing the *Pnma* phase with respect to the others. We fully characterize the *Pnma* ground-state structure, computing the full phonon dispersion curves and assigning the Raman and IR zone-center modes systematically. The mechanical properties are also presented, and the calculated heat capacity for orthorhombic $SrRuO_3$ is compared with experimental data.

The remainder of the paper is organized as follows. In Section 2, we describe the technical details of the first-principles calculations used in this work. Section 3 presents the lattice parameters, electronic structure, magnetic moment, phonon dispersion curves and mechanical properties of cubic $SrRuO_3$. The antiferrodistortive phonon instabilities, structural properties and the relative internal energies of different phases are discussed in Section 4. Then we present the zone-center phonons, mechanical properties and specific heat for the ground state of $SrRuO_3$ in Section 5. Finally, the conclusions are given in Section 6.

2. Methodologies

Our calculations were based on density functional theory (DFT) using a plane-wave basis set as implemented in the ABINIT package [29]. Optimized norm-conserving pseudopotentials [30] including scalar-relativistic effects [31] were generated with the OPIUM code [32]. Non-local pseudopotentials [33] were designed for Sr and Ru to enhance transferability. The 2$s$ and 2$p$ levels of O and 4$s$, 4$p$, 4$d$ and 5$s$ levels of Sr/Ru were treated as valence electrons. All calculations were spin-polarized and have been performed with the ferromagnetic (FM) solution.

Since most recent theoretical and experimental works suggest that SrRuO$_3$ is a weakly correlated compound whose physical properties can be well described in the frame work of standard DFT [27,28], we considered both the Wu-Cohen generalized gradient approximation (WC-GGA) [34] and the local spin density approximation (LSDA) [35] as exchange-correlation functionals. As it will appear in the next sections, the structural properties of SrRuO$_3$ were better reproduced by WC-GGA than LSDA, so that, herein the WC-GGA functional was adopted for all the phonon calculations unless stated otherwise.

Convergence was reached for an energy cut-off of 30 hartree for the plane-wave expansion and a 16×16×16 grid of $k$-points for the Brillouin zone sampling of the 5-atom cubic perovskite cell. For a 20-atom cell, which contains 4 formula units (f.u.) and corresponds to a $\sqrt{2}\times\sqrt{2}\times2$ repetition of the 5-atom cubic cell, a $k$-mesh of 8×8×6 was used. We adopted a 6×6×6 $k$-grid for the phases with a 40-atom cell (2×2×2 repetition of the 5-atom cell). Structural optimizations were performed independently with each functional until the difference of forces (resp. stresses) were smaller than 5×10$^{-5}$ hartree/Bohr (resp. 5×10$^{-7}$ hartree/Bohr$^3$). Wave functions were converged until the difference of the total energy was smaller than 10$^{-11}$ hartree for the cubic phase and until the difference of the forces was smaller than 5×10$^{-6}$ hartree/Bohr for distorted phases. The phonon frequencies and elastic constants were computed according to density functional perturbation theory (DFPT) [36,37].

The CRYSTAL09 code [38] (with the WC-GGA functional and the B1-WC [39] hybrid functional) has also been used to check the consistency of the results from ABINIT. These calculations were converged until the difference on the total energy is smaller than $10^{-8}$ hartree. The structures were relaxed until the root mean squarte of the gradient and displacements are smaller than $1.5\times10^{-4}$ and $1.2\times10^{-3}$, respectively.

3.  Cubic Perovskite Structure

First we considered the high-symmetry cubic perovskite structure ($Pm\bar{3}m$) of SrRuO$_3$. The results after full structural relaxation are summarized in Table 1, together with available experimental and other theoretical data. Within the WC-GGA, our calculated lattice constant (3.930 Å) is in good agreement with experiment [8], confirming the accuracy of this functional. A slight underestimation of the experimental lattice constant, typical of the LSDA functional, can also be observed, however, our calculated value of 3.888Å is consistent with previous theoretical work [25]. For the calculated magnetic moment, the present result from LSDA also agrees well with previous theoretical values in the literature [16].

In Figure 1, we present the calculated electronic band structure and density of states (DOS) for cubic SrRuO$_3$ within the WC-GGA. Very similar results have been obtained from LSDA. In our calculation, cubic SrRuO$_3$ appears to be metallic. An excellent agreement can be observed from the electronic band structure and DOS compared with the previous LSDA studies of Singh [16] and Allen *et al* [17]. Our GGA results are in line with the recent proposal that GGA is appropriate for SrRuO$_3$ [27], indicating that the electronic structure feature of SrRuO$_3$ can be well captured by WC-GGA functional, and also confirming the validity of the generated pseudopotentials and the other parameters used in this work.

The phonon dispersion curves and the phonon DOS of cubic SrRuO$_3$ are plotted in Figure 2. In the dispersion curves (left panel in Figure 2), the $\Gamma$-$X$, $\Gamma$-$M$, and $\Gamma$-$R$ lines are along the [100], [110], and [111] directions in reciprocal space, respectively. Imaginary frequencies indicate dynamical instabilities of the crystals. No unstable mode at the $\Gamma$ point is observed while strongly unstable antiferrodistortive (AFD) phonon modes appear at the $M$ ($M_3^+$ mode, ~211$i$ cm$^{-1}$) and the $R$ ($R_4^+$ mode, ~223$i$ cm$^{-1}$) high-symmetry points as well as along the lines joining them (edges of the Brillouin zone). The $M_3^+$ and $R_4^+$ unstable modes are associated to pure rotations of oxygen octahedra in which consecutive octahedra along the rotation axis rotate either in-phase ($M_3^+$ mode) or anti-phase ($R_4^+$ mode) as illustrated in Figure 3. This is consistent with the projected phonon DOS (right panel in Figure 2), where the unstable region corresponds to pure oxygen motions.

Although AFD unstable modes are common in $ABO_3$ cubic perovskites, the amplitudes of instability for $SrRuO_3$ are somewhat surprising, in view of its Goldschmidt tolerance factor [40-43] close to 1 ($t$=0.994), suggesting that the atoms nearly perfectly fit with the cubic perovskite structure. We notice, however, that in spite of their tolerance factor equal or close to 1, AFD instabilities were also reported in $BaZrO_3$ [44] ($t$=1.004) and $SrTiO_3$ [45] ($t$=1.002). To check the reliability of the present calculations, we also performed the phonon calculation with LSDA in ABINIT, and with WC-GGA and the B1-WC [39] hybrid functional using the CRYSTAL09 code [38]. In all cases, we reproduced the instabilities at the $R$ and $M$ points, attesting that the present results are independent of exchange-correlation functionals or codes.

Finally, the elastic stiffness constants of cubic $SrRuO_3$ have been computed within WC-GGA from the strain DFPT method [37]. For cubic crystals, there are 3 independent elastic constants, i.e., $c_{11}$, $c_{12}$, $c_{44}$. The calculated values are 312.8 GPa, 101.8 GPa and 65.4 GPa, respectively. According to the Voigt–Reuss–Hill (VRH) approximations [46-48], the bulk modulus $B=(c_{11}+2c_{12})/3$ of cubic $SrRuO_3$ is estimated to be 172 GPa. No experimental data is available for comparison, but the bulk modulus of 200 GPa and 219 GPa have been reported in previous LSDA calculations by fitting to the equation of states [25], which likely overestimates $B$ due to the typical overbinding tendency of LSDA.

4. Antiferrodistortive Instabilities and Distorted Phases

As previously discussed, the cubic structure of SrRuO$_3$ exhibits strong antiferrodistortive instabilities, associated with rotations of the oxygen octahedra. If, in line with Glazer [49], we restrict ourselves to the oxygen rotations associated with $M_3^+$ and $R_4^+$ modes and consider that such rotations can appear along any of the three cubic directions, this defines 6 distinct basic tilt patterns (spanning the 6-dimensional $M_3^+ \oplus R_4^+$ reducible representation of the $Pm\bar{3}m$ phase) that can be combined to generate various tilted structures. Using group theory analysis, Howard and Stokes [50] demonstrated the existence of 15 distinct combinations of basic tilt patterns that, when condensed within the $Pm\bar{3}m$ phase, will lower the symmetry to distinct subgroups. The resulting oxygen-tilted structures are usually specified using compact Glazer's notations [49] like $a^0b^+c^-$ in which the three literals refer to the 3 cubic directions and the 0, + or – superscripts refer to the condensation of no tilt, $M_3^+$ tilt or $R_4^+$ tilt along each specific direction respectively (the use of the same letter along two directions indicates tilts of same amplitude).

In this context, the observed sequence of structural phase transitions of SrRuO$_3$ can be understood as a successive appearance of different unstable tilt patterns in the cubic structure:

$$Pm\bar{3}m\ (a^0a^0a^0) \rightarrow I4/mcm\ (a^0a^0c^-) \rightarrow [\ Imma\ (a^0b^-b^-)\ ] \rightarrow Pnma\ (a^-b^+a^-)$$

But why does this material evolve to a specific *Pnma* ground state structure? The strengths of the $M_3^+$ and $R_4^+$ instabilities in the cubic structure are almost equal and it is not a priori trivial to understand why a specific combination of tilts is preferred over others. It is worth noticing that the structures of the various possible tilted phases are not restricted to $M_3^+$ and $R_4^+$ octahedra rotations: for a specific combination of tilts, the condensation of a given tilt pattern will lower the symmetry to a certain space group within which the system will further relax through anharmonic couplings with other modes that might further stabilize that phase.

Distortions in cubic perovskites are usually understood as a way to improve cationic coordinations and are rationalized in term of atomic radii through the empirical Goldschmidt tolrance factor, *t* [40]. In fact, most cubic perovskites (typically with *t* <

1, *i.e.* in which the A cation is small and under-coordinated) exhibit a *Pnma* tilted ground state [41]. Thomas [13] and Woodwards [14] reported that anti-polar A-cation motions, allowed by symmetry in the *Pnma* phase, can play an important role in improving A-atom coordination and contribute to stabilize that phase over other distorted structures. This was recently highlighted at the first-principles level by Benedek and Fennie [15], from the study of a series of insulating $ABO_3$ compounds with t<1. It is not obvious if these latter results generalise to $SrRuO_3$ which is metallic and exhibits a tolerance factor very close to 1 ($t = 0.994$). In order to provide a more comprehensive and quantitative understanding on the origin of the *Pnma* ground-state structure of $SrRuO_3$, we propose below an original decomposition of the energy gain produced independently from oxygen and cationic motions in many metastable phases.

We performed systematic structural relaxation calculations of various tilted systems including one, two or three distinct tilts. The results are reported in Figure 4 (also see Table 6 and Table 7 in the Appendix), where we decompose the total energy gain from the undistorted cubic structure to each tilted system in terms of the contributions $E_{\text{oxygen}}$, $E_{\text{cation}}$ and $E_{\text{strain}}$ (as defined in the caption of Figure 4). In the following we will attempt to rationalize the relative stability of each tilted system by considering each of these contributions.

First we notice that, out of the various tilt systems, the $a^-b^+a^-$ phase indeed produces the largest gain of energy, which corresponds to the *Pnma* ground state of $SrRuO_3$ as seen in experiments. The calculated relative energy for the fully relaxed *Pnma* phase is -219 meV/f.u using WC-GGA (-201.6 meV/f.u from LSDA), comparable with the previous LSDA calculations of -188 meV/f.u. by VASP, while slightly larger than the value of -150 meV/f.u. by SIESTA [25] as well as -140 meV/f.u. by LAPW [16].

The next immediate observation from Figure 4 is that the appearance of the first rotation provides the largest gain of energy – the condensation of a second or third rotation in a different axis does not proportionally increase the gain of energy, due to a positive biquadratic (competitive) coupling between them. It is also immediately clear that out of all of the energy contributions, $E_{\text{oxygen}}$, which is related to $M_3^+$ and $R_4^+$ oxygen motions [51], dominates the total gain of energy – antipolar motions and strain relaxation (related to the contributions $E_{\text{cation}}$ and $E_{\text{strain}}$ respectively) have a smaller effect. However, whilst all these contributions (from second and third rotations, strain and antipolar motions) are smaller, they are of the scale of the relative energy difference between phases, i.e., their contributions are crucial in determining the ground state, as we will explore next.

When looking more closely at the contribution of the energy gain from only oxygen motion ($E_{oxygen}$) across all the phases it is apparent that the combination of two "−" rotations is the most favorable. In other words, if the structures could only allow for oxygen rotations, the $a^0b^-b^-$, $a^-b^+a^-$ and $a^-a^-a^-$ phases would be nearly at the same energy, as a third "+" or "−" rotation does not noticeably reduce the energy further.

When including oxygen atomic relaxation and strontium motions, the $a^0b^+c^-$, $a^-b^+a^-$ and $a^+a^+c^-$ phases obtain the greatest additional energy gain ($E_{cation}$). To understand the origin of this energy gain from $E_{cation}$, we present in Table 2 an analysis of the atomic distortions in terms of contributions from different symmetry-adapted modes [52,53]. Relevant invariant polynomials coupling these modes in the Landau-type expansion of the energy around the cubic phase [54] for various tilt systems are reported in Appendix (Table 8). As already mentioned the $R_4^+$ and $M_3^+$ modes correspond to the oxygen octahedra rotations, while the $X_5^+$ and $R_5^+$ correspond to Sr (and to a lesser degree oxygen) anti-polar distortions; $M_2^+$ and $M_4^+$ are associated to other purely oxygen motions. See Figure 3 for a schematic illustration of these distortions. Among investigated phases, the *Pnma* phase exhibits the largest total anti-polar distortion, especially from the $X_5^+$ mode. $X_5^+$ is not by itself an unstable mode as can be seen from the cubic phonon dispersion curves (see Figure 2), and its appearance is a signature of a "hybrid" improper-type behavior [55,56] (i.e. it appears through a mechanism similar to "hybrid" improper ferroelectricity in artificial superlattices [55]). It is the trilinear coupling term (see Appendix Table 8), $Q(R_4^+)Q(M_3^+)Q(X_5^+)$ [57], that drives the appearance of $X_5^+$ motion. This trilinear term can only be introduced by the coexistence of an in-phase rotation (+) and an out-of-phase rotation (−) (as in the $a^0b^+c^-$, $a^-b^+a^-$ and $a^+a^+c^-$ phases). Neither the combination of two out-of-phase rotations ($a^0b^-b^-$) nor two in-phase rotations ($a^0b^+b^+$) introduces a trilinear coupling, although it can induce small anti-polar $R_5^+$ distortions and $M_4^+$ oxygen motions through bi-couplings of the form $Q(M_4^+)Q(M_3^+)^2$ and $Q(R_4^+)^3Q(R_5^+)$. It is therefore the combination of a "+" and "−" rotation allowing for a trilinear coupling with the $X_5^+$ anti-polar mode that necessarily lowers the energy of the ground state structure in metallic SrRuO$_3$.

We note that the condensation of the anti-polar modes affect the magnitude of $M_3^+$ and $R_4^+$. Once the Sr motion is included, the oxygen rotation angles are enhanced as can be seen in Appendix Table 6. This cooperative coupling is enabled through the trilinear term. With the anti-polar motion the energy gain produced by the pure oxygen motion [58] is reduced (as shown by the dotted line in Figure 4) since the rotation magnitude is no longer at the minimum of the double well. This effect is particularly strong in the *Pnma* phase which is consistent with the strong coupling with the $X_5^+$ mode.

Finally, strain relaxation (related to the contribution $E_{strain}$) appears to play the largest role in the $a^0a^0c^-$, $a^0a^0c^+$ and $a^-a^-a^-$ phases, which interestingly are the only phases that do not exhibit anti-polar distortions. Within the remaining phases, the couplings with anti-polar distortions (related to the contribution $E_{cation}$) play a more significant role in stabilizing structures than the strain effect.

Therefore, in summary, we suggest that the $a^-b^+a^-$ ground state of SrRuO$_3$ takes advantage of the features of both the $a^0b^-b^-$ and $a^0b^+c^-$ phases, i.e. the combination of two "−" rotations which appears to be the most favorable oxygen motion and the combination of a "+" and "−" rotation to induce the *X* point anti-polar mode through a trilinear coupling.

The calculated structural parameters for several possible intermediate phases of SrRuO$_3$ are collected in Table 3. The Wyckoff positions are obtained from the fully relaxed structure using the FINDSYM code [59]. All these phases have a very similar lattice volume of approximately 241 Å$^3$ per 20-atom cell. The calculated lattice parameters and atomic positions for the *Pnma* phase are in fair agreement with the available experiment data [12,60-62] and the theoretical results (see Table 9 in Appendix). The calculated magnetic moments of 1.98 $\mu_B$/f.u from WC-GGA and 1.70 $\mu_B$/f.u from LSDA are in reasonable agreement with the previously reported theoretical and experimental data. The "effective magnetic moment" of SrRuO$_3$ can be estimated as $\mu_{eff} = [\mu_B(\mu_B+g_0)]^{0.5}$ (with $g_0 = 2$) using a simple spin-only magnetism approach [18]. Here we obtain a $\mu_{eff}$ of 2.81$\mu_B$ from WC-GGA and 2.51$\mu_B$ from LSDA, which are also comparable to the experimental values of 2.4$\mu_B$-2.83$\mu_B$ [19,63]. It is worth mentioning that without including self-correction or Hubbard U for the on-site correlations, our calculations reproduce the structural properties and magnetic moments for SrRuO$_3$ reasonably, indicating that WC-GGA is a favorable approach for this weakly correlated material. For the *I4/mcm* phase, our calculated lattice volume is slightly smaller (~2.4%) than the experiment, which could be attributed to the fact that the experimental data were gathered at high temperature (823 K) [8]. For the proposed *P4/mbm* phase, the calculated lattice parameters and magnetic moment are very close to the *I4/mcm*, but with a higher energy. Note that the calculated *Imma* and *Cmcm* phases show rather different structural properties while being very close in magnetic moment of 1.87$\mu_B$/f.u. and energies of ~196 meV.

The rotation angles listed in Table 3 (also see Figure 8 and Table 6 in Appendix) are calculated from the amplitudes of the overlaps of the atomic distortions with rigid $R_4^+$ and $M_3^+$ motions along three Cartesian directions. Though our calculated atomic positions and lattice parameters of SrRuO$_3$ are in reasonable agreement with the experiments, the rotation angles are overestimated. The amplitude of individual tilts of RuO$_6$ decreases with the condensation of additional antiferrodistortive instabilities, from cubic to the orthorhombic phase. For example, *I4/mcm* ($a^0b^0c^-$) → *Imma* ($a^0b^-b^-$) → *Pnma* ($a^-b^+a^-$), the out of phase rotation angle of *I4/mcm* is 11.53°, which decreases to 8.49° with the appearance of a second $R_4^+$ tilt, and is further reduced (7.34°) by another $M_3^+$ tilt, resulting in the *Pnma* ($a^-b^+a^-$) ground state. Similar trends can be observed for the other phases, revealing that the newly introduced rotation is interacting with the previous one(s). This confirms the previous observation that rotations compete with each other.

5. Orthorhombic SrRuO$_3$

In this section, we further characterize the *Pnma* structure of SrRuO$_3$. The full phonon dispersion curves and the phonon DOS are reported in Figure 5. No imaginary frequency is observed, in agreement with the fact that this phase is the ground state. We can decompose the phonon DOS in three consecutive regions: the lowest part of the spectrum (around 100 cm$^{-1}$) is dominated by Sr motions, then an intermediate range (100-300 cm$^{-1}$) is associated to Ru and O motions, while the upper part (above 300 cm$^{-1}$) is associated to quasi purely O motion and is made of two regions separated by a gap between 400 and 500 cm$^{-1}$. Comparing with the DOS of the cubic structure (Figure 1), we see strong similarities except that the unstable branches of the cubic phase have been stabilized in the *Pnma* phase, producing the oxygen bands in the 300-400 cm$^{-1}$ range.

For perovskite-like ABO$_3$ compounds with a *Pnma* structure, there are 60 $\Gamma$-phonon frequencies in total. The irreducible representations of orthorhombic SrRuO$_3$ at the $\Gamma$ point are: 7A$_g$ $\oplus$ 5B$_{1g}$ $\oplus$ 7B$_{2g}$ $\oplus$ 5B$_{3g}$ $\oplus$ 10B$_{1u}$ $\oplus$ 8B$_{2u}$ $\oplus$ 10B$_{3u}$ $\oplus$ 8A$_u$, among which, 3 modes are acoustic (B$_{1u}$, B$_{2u}$, B$_{3u}$), 8 are silent (A$_u$), 24 are Raman active (A$_g$, B$_{1g}$, B$_{2g}$ and B$_{3g}$) and the remaining 25 modes are infrared (IR) active (B$_{1u}$, B$_{2u}$ and B$_{3u}$). The 8 A$_u$ silent modes, which will not be discussed further, are calculated at frequencies of 97, 141, 181, 228, 280, 342, 540 and 546 cm$^{-1}$.

The calculated Raman active mode frequencies of orthorhombic SrRuO$_3$ are given in Table 4 and compared with available experimental data which were taken from measurements on SrRuO$_3$ thin films [64-68]. Note that, the Raman modes measured by Kirillov *et al* [64] and Tenne *et al* [66] which were not assigned in their experiments are also shown in Table 4 and sorted by quantity for comparison with our calculations. It can be observed that our assignments and frequencies for the *Pnma* phase of SrRuO$_3$ are in an overall good agreement with experimental measurements [64-68]. The large difference between the calculated values of A$_g$ (174 cm$^{-1}$) and B$_{2g}$ (349 cm$^{-1}$) modes and the experimental data of A$_g$ (225 cm$^{-1}$) and B$_{2g}$ (390 cm$^{-1}$) modes from Iliev *et al* [65] correlates with the fact that their measured Raman frequencies are generally higher than the other experiments as observed in the table. This may be attributed to the substrate influence (e.g., the epitaxial constraint imposed by the substrate) on the vibrational properties of SrRuO$_3$ thin films or other possible different experimental conditions. Also, in their report, it was proposed that the frequency of the first $A_g$ mode was 123 cm$^{-1}$, much larger than our calculated value of 98 cm$^{-1}$ and the other experimental data of 94 cm$^{-1}$ or 98 cm$^{-1}$, but very close to the next lowest frequency modes at 131 cm$^{-1}$, 118 cm$^{-1}$ and 130 cm$^{-1}$ as listed in Table 3. Hence, we argue that the first measured Raman mode in Iliev *et al.*'s work could correspond to the second lowest frequency mode in our calculation and the other experiments. For the $A_g$ modes, the highest one at 612 cm$^{-1}$ was not detected by experiments possibly because the experimental measurements are often limited to frequencies below 600 cm$^{-1}$. Among the $B_{1g}$, $B_{2g}$ and $B_{3g}$ modes, only two of the $B_{2g}$ modes were reported by experiments, where the frequencies are consistent with our calculations within 10 cm$^{-1}$. We believe our calculated results can serve as a reference for future theoretical or experimental research on Raman spectroscopy.

In Table 5, we report the calculated IR frequencies of the transverse optic phonon modes. Since the charges of the IR phonons are screened by free carriers in metallic SrRuO$_3$, it is difficult to observe all the IR modes experimentally. As for IR spectra, there is no experimental measurement on bulk orthorhombic SrRuO$_3$ single crystals. We compare our results with the IR frequencies of SrRuO$_3$ thin films fabricated on [0 0 1] oriented SrTiO$_3$ substrates using pulsed laser deposition [69]. Only seven IR active modes appeared at 300 K in the measurements of Ref. [69] as shown in Table 5. In their experiments, the assignment of modes was supported by shell model calculations of the lattice dynamics of SrRuO$_3$. Using a shell model to support the assignment of experimentally observed Raman and IR peaks can be reasonably good but not always accurate as suggested by our previous work on a related compound [70]. Here, according to our first-principles results, the two experimentally observed modes with frequencies of 380 cm$^{-1}$ and 435 cm$^{-1}$, which were assigned to be $B_{1u}$ modes by the shell model calculations [69], would be the $B_{3u}$ modes. These are much closer to our calculated values of 382 cm$^{-1}$ and 424 cm$^{-1}$, while the experimental $B_{3u}$ mode of 328 cm$^{-1}$ may correspond to the $B_{1u}$ mode with a frequency of 318 cm$^{-1}$ in our calculation. After this reassignment of the IR modes, the agreement between our results and experimental data is rather good. Note that Crandles *et al* [71] also detected a few IR modes in their experiments, with frequencies of 150 cm$^{-1}$, 300cm$^{-1}$ and 552 cm$^{-1}$, which is consistent with our calculated values of $B_{1u}$ of 159 cm$^{-1}$, 318 cm$^{-1}$ and 555 cm$^{-1}$. However, these modes were not assigned in their work. We hope to encourage future experimental study of the IR and Raman spectra of single crystal SrRuO$_3$ to compare with the present calculated results.

From the strained DFPT calculations, the elastic constants for orthorhombic SrRuO$_3$ were obtained with all the ions relaxed in the cell. The calculated values are 272.0, 263.8, 323.2, 73.0, 67.4, 94.3, 162.9, 130.5, and 110.0 GPa for $c_{11}$, $c_{22}$, $c_{33}$, $c_{44}$, $c_{55}$, $c_{66}$, $c_{12}$, $c_{13}$, and $c_{23}$, respectively. The bulk (*B*), shear (*G*) and Young's (*E*) modulus can be derived according to the VRH approximation [46-48] from these elastic constants. The calculated bulk modulus is 184.8 GPa, in good agreement with the experimental values of 192(3) GPa [72] and 180 GPa~190 GPa [73], as well as the theoretical value of 190.5 GPa [74], but larger than the value of 171.5 GPa from Yamanaka *et al.*'s experiment [75]. The shear modulus and Young's modulus are 75.2 GPa and 198.7 GPa, still slightly larger than the values of 60.1 GPa and 161 GPa in Reference [75], respectively. Using $H_V=G^2/(3B+G)$ [76], the micro hardness ($H_V$) of SrRuO$_3$ is estimated to be 9.0 GPa, smaller than the experimental value of 12.7 GPa [75]. Furthermore, we deduced the Debye temperature $\Theta_D$ from the longitudinal ($v_l$) and transverse ($v_t$) elastic wave velocities. The calculated $v_l$ and $v_t$ are 3396 and 6612 m/s, comparable to experimental data of 3083 and 6312 m/s [75]. Our estimated $\Theta_D$ for orthorhombic SrRuO$_3$ is 493.9 K, which is in reasonable agreement with the experimental results of 525.5 K [77] and 448 K [75], but much larger than that of 368 K from Allen *et al.*'s work [17], and smaller than that of 601(41) K from Chakoumakos *et al.*'s experiment [10], respectively. The wide spread of experimental $\Theta_D$ can be related to different issues. On one hand, $\Theta_D$ was extracted from the fit of different measured quantities (sound velocity, volume or specific heat) in different temperature ranges. On the other hand, the measurements were carried out on polycrystalline samples of SrRuO$_3$ with varied quality and different experimental conditions.

The lattice contribution to the constant volume heat capacity (C$_V$) was calculated for orthorhombic SrRuO$_3$ using the method introduced by Lee and Gonze [78], and compared with available experimental results in Figure 6. In metals like SrRuO$_3$, there is an additional electronic contribution to C$_V$ [17], however, except in the vicinity of 0K, it is negligible in comparison to the lattice one and not included in the present calculation. In the whole range of stability of the *Pnma* phase, the calculated C$_V$ is in excellent agreement with the experiment [17]. Only in the higher temperature range (≥300 K), our calculation starts deviating from the experiment [75] as can be expected due to the anharmonic effects not included in our calculation. Although in absence of experimental phonon measurement outside $\Gamma$, this excellent agreement obtained without adjustable parameters indeed validates the reliability of our phonon dispersion curves.

Finally, the calculated electronic DOS is plotted in Figure 7. From our WC-GGA calculation, the ground state $SrRuO_3$ is also metallic and in the limit of a half-metal. It is comparable with pseudo-SIC calculation [25] and LSDA results [16,18,79], where the majority spin exhibits a very small electronic DOS at the Fermi level. We note that the features of the LSDA electronic DOS from Rondinelli *et al.*'s work [25], especially near the Fermi level, are slightly different from ours as well as some other works [16,18,79]. For the majority spins, the Fermi level is located in the central part of the peak of the total DOS in their work [25], while it is near the bottom in our calculation as well as other studies [16,18,79] (*i.e.*, the spin splitting is smaller in the work of Rondinelli *et al.* as it also appears in their smaller magnetic moment). Calculations with the GW approximation predicted instead that there is a band gap in the spin-up channel at the Fermi level [79], suggesting that orthorhombic $SrRuO_3$ is a half metal, but this has not been confirmed experimentally.

6. Conclusions

In this paper, we have presented a systematic first-principles study of various properties of SrRuO$_3$. The calculated lattice parameters, magnetic moments and mechanical properties are in good agreement with the available theoretical and experimental data. We mainly focused on the dynamical properties of cubic and orthorhombic phases which, to the best of our knowledge, have not been computed before from first-principles. Strong antiferrodistortive instabilities in cubic SrRuO$_3$ were found, which drive the system towards its *Pnma* ground state. We clarified that the coexistence of "−" and "−" tilts as well as "+" and "−" tilts in the *Pnma* ($a^-b^+a^-$) phase of SrRuO$_3$ account for the favorability of this ground state respect to other distorted structures: the combination of two "−" rotations appears to be the most favorable oxygen motion while the combination of a "+" and "−" rotation induces the appearance of an anti-polar $X_5^+$ mode through a trilinear coupling, further reducing the total energy of the structure. Our original analysis quantifies at the first-principles level the important role of A-cation motions in stabilizing the *Pnma* ground state of SrRuO$_3$ and complements the recent work of Benedek and Fennie [15]. The lattice dynamical properties of the *Pnma* phase of SrRuO$_3$ have been fully characterized. We proposed partial reassignment of experimental data at the Brillouin zone-center. The full dispersion curves have also been obtained, constituting benchmark results for the interpretation of future measurements and providing access to thermodynamical properties that had never been computed before.

ACKNOWLEDGEMENT

This work is supported by the ARC project TheMoTherm (Grant No. 10/15-03). N. M. would like to thank Daniel Bilc for guidance in using CRYSTALl09 code and Eric J. Walter for valuable discussions on the pseudopotentials generation as well as Eric Bousquet for help with figure production. P. G. thanks the Francqui Foundation for a Research Professorship. The CECI (http://www.ceci-hpc.be, F.R.S.-FNRS Grant 2.5020.11), PRACE-2IP on Huygens and Hector (EU FP7 Grant RI-283493), CISM-UCLouvain, and SEGI-ULg are greatly acknowledged for providing computational resources.

APPENDIX

A. Definition of rotation angle

In the literature, the rotation angles in perovskites are defined in different ways [20,49] or calculated by different empirical equations [8,80]. However, for complex tilt patterns, it is not always easy to work out the contributions from individual rotations with the previous methods. Here, the rotation angles are calculated as follows: First, the amplitude $A$ of each $M_3^+$ or $R_4^+$ mode within the structure is obtained from an overlap of the atomic distortion with the relevant phonon eigenvectors. Then the amplitude is converted into a rotation angle as illustrated in Figure 8. The calculated rotations angles are presented in Table 6.

B. The gain of energy and energy expansion

We report in Table 7 the values of the gains of energy for different phases of $SrRuO_3$ as plotted in Figure 4. In Table 8, we collect relevant invariant polynomials [54] of the Landau-type energy expansion around the cubic phase. All these tables have been discussed in the main text. For more details and discussion, please refer to Section 4.

C. The low-temperature ground state structure

The structural properties of ground state $SrRuO_3$ are collected in Table 9. It can be observed that the calculated lattice parameters, atomic positions and magnetic moments of the *Pnma* $SrRuO_3$ phase are in good agreement with previous theoretical and experimental data.

Reference

[1] P. Zubko, S. Gariglio, M. Gabay, et al. *Annu. Rev. Condens. Matter Phys.* 2012 **2** 141
[2] H. Hwang, Y. Iwasa, M. Kawasaki, et al. *Nature Mater.* 2012 **11** 103
[3] J. M. Rondinelli and N. A. Spaldin *Adv. Mater.* 2011 **23** 3363
[4] J. Junquera and P. Ghosez *Nature* 2003 **422** 506
[5] M. Stengel and N. A. Spaldin *Nature* 2006 **443** 679
[6] M. Verissimo-Alves, P. García-Fernández, D. I. Bilc, et al. *Phys. Rev. Lett.* 2012 **108** 107003
[7] G. Koster, L. Klein, W. Siemons, et al. *Rev. Mod. Phys.* 2012 **84** 253
[8] B. J. Kennedy and B. A. Hunter *Phys. Rev. B* 1998 **58** 653
[9] B. J. Kennedy, B. A. Hunter, and J. R. Hester *Phys. Rev. B* 2002 **65** 224103
[10] B. Chakoumakos, S. Nagler, S. Misture, et al. *Physica B* 1998 **241** 358
[11] S. Cuffini, J. Guevara, and Y. Mascarenhas, in *Mater. Sci. Forum* (Trans Tech Publ, 1996), p. 789.
[12] C. Jones, P. Battle, P. Lightfoot, et al. *Acta Cryst. Sect. C* 1989 **45** 365
[13] N. W. Thomas *Acta Cryst. Sect. B* 1996 **52** 16
[14] P. M. Woodward *Acta Cryst. Sect. B* 1997 **53** 44
[15] N. A. Benedek and C. J. Fennie *J. Phys. Chem. C* 2013 **117** 13339
[16] D. J. Singh *J. Appl. Phys.* 1996 **79** 4818
[17] P. Allen, H. Berger, O. Chauvet, et al. *Phys. Rev. B* 1996 **53** 4393
[18] G. Santi and T. Jarlborg *J. Phys.: Condens. Matter* 1997 **9** 9563
[19] M. Shepard, S. McCall, G. Cao, et al. *J. Appl. Phys.* 1997 **81** 4978
[20] A. Zayak, X. Huang, J. Neaton, et al. *Phys. Rev. B* 2006 **74** 094104
[21] A. Zayak, X. Huang, J. Neaton, et al. *Phys. Rev. B* 2008 **77** 214410
[22] K. J. Choi, S. H. Baek, H. W. Jang, et al. *Adv. Mater.* 2010 **22** 759
[23] D. Kan and Y. Shimakawa *Cryst. Growth Des.* 2012 **11** 5483
[24] D. Toyota, I. Ohkubo, H. Kumigashira, et al. *Appl. Phys. Lett.* 2005 **87** 162508
[25] J. M. Rondinelli, N. M. Caffrey, S. Sanvito, et al. *Phys. Rev. B* 2008 **78** 155107
[26] P. García-Fernández, D. I. Bilc, P. Ghosez, et al. *Phys. Rev. B* 2012 **86** 085305
[27] C. Etz, I. Maznichenko, D. Böttcher, et al. *Phys. Rev. B* 2012 **86** 064441
[28] S. Grebinskij, S. Masys, S. Mickevicius, et al. *Phys. Rev. B* 2013 **87** 035106
[29] X. Gonze, J. M. Beuken, R. Caracas, et al. *Comput. Mater. Sci.* 2002 **25** 478
[30] A. M. Rappe, K. M. Rabe, E. Kaxiras, et al. *Phys. Rev. B* 1990 **41** 1227
[31] I. Grinberg, N. J. Ramer, and A. M. Rappe *Phys. Rev. B* 2000 **62** 2311
[32] OPIUM pseudopotential package, *http://opium.sourceforge.net*
[33] N. J. Ramer and A. M. Rappe *Phys. Rev. B* 1999 **59** 12471
[34] Z. Wu and R. E. Cohen *Phys. Rev. B* 2006 **73** 235116
[35] J. P. Perdew and Y. Wang *Phys. Rev. B* 1992 **45** 13244
[36] X. Gonze and C. Lee *Phys. Rev. B* 1997 **55** 10355
[37] D. Hamann, X. Wu, K. M. Rabe, et al. *Phys. Rev. B* 2005 **71** 035117
[38] R. Dovesi, R. Orlando, B. Civalleri, et al. *Z. Kristallogr.* 2005 **220** 571
[39] D. Bilc, R. Orlando, R. Shaltaf, et al. *Phys. Rev. B* 2008 **77** 165107
[40] V. M. Goldschmidt, T. F. Barth, G. Lunde, et al. *Oslo, Mat.-Nat. Kl. 2, 117* 1926
[41] M. W. Lufaso and P. M. Woodward *Acta Cryst. Sect. B* 2001 **57** 725
[42] Calculated from Shannon Radii, R. D. Shannon, *Acta Cryst. Sect. A 1976* **32**, *751*
[43] Generally, the value of *t* obtained here are closed to the results calculated from bond valance model. Both methods has been implemented in the SPuDs code.
[44] J. W. Bennett, I. Grinberg, and A. M. Rappe *Phys. Rev. B* 2006 **73** 180102
[45] C. Lasota, C.-Z. Wang, R. Yu, et al. *Ferroelectrics* 1997 **194** 109
[46] W. Voigt *Lehrbuch der Kristallphysik, B.G. Teubner, Leipzig.* 1928
[47] A. Reuss *Z. Angew. Math. Mech.* 1929 **9** 49
[48] R. Hill *Proc. Phys. Soc. London* 1952 **65** 349
[49] A. Glazer *Acta Cryst. Sect. B* 1972 **28** 3384
[50] C. J. Howard and H. Stokes *Acta Cryst. Sect. B* 1998 **54** 782
[51] In some phases (see Table 2) additional oxygen motions ($M_2^+$ and $M_4^+$) are allowed by symmetry. However in practice the amplitude remains negligible.
[52] D. Orobengoa, C. Capillas, M. I. Aroyo, et al. *J. Appl. Cryst.* 2009 **42** 820

[53] J. Perez-Mato, D. Orobengoa, and M. Aroyo *Acta Cryst. Sect. A* 2010 **66** 558
[54] H. T. Stokes and D. M. Hatch INVARIANTS, *www.physics.byu.edu/~stokesh/isotropy.html*
[55] E. Bousquet, M. Dawber, N. Stucki, et al. *Nature* 2008 **452** 732
[56] N. A. Benedek and C. J. Fennie *Phys. Rev. Lett.* 2011 **106** 107204
[57] S. Amisi, E. Bousquet, K. Katcho, et al. *Phys. Rev. B* 2012 **85** 064112
[58] Terms involving only $R_4^+$ and $M_3^+$ in the Landau-type expansion.
[59] H. T. Stokes, B. J. Campbell, and D. M. Hatch FINDSYM, *stokes.byu.edu/isotropy.html*. 2013
[60] J. Longo, P. Raccah, and J. Goodenough *J. Appl. Phys.* 1968 **39** 1327
[61] A. Kanbayasi *J. Phys. Soc. Jpn.* 1976 **41** 1876
[62] S. Bushmeleva, V. Y. Pomjakushin, E. Pomjakushina, et al. *J. Magn. Magn. Mater.* 2006 **305** 491
[63] A. Kanbayasi *J. Phys. Soc. Jpn.* 1978 **44** 108
[64] D. Kirillov, Y. Suzuki, L. Antognazza, et al. *Phys. Rev. B* 1995 **511** 75813
[65] M. Iliev, A. Litvinchuk, H. G. Lee, et al. *Phys. Rev. B* 1999 **59** 364
[66] D. A. Tenne and X. Xi *J. Am. Ceram. Soc.* 2008 **91** 1820
[67] G. Herranz, F. Sánchez, J. Fontcuberta, et al. *Phys. Rev. B* 2005 **71** 174411
[68] S. B. Anooz, J. Schwarzkopf, R. Dirsyte, et al. *Phys. Status Solidi a* 2010 **207** 2492
[69] D. Crandles, F. Eftekhari, R. Faust, et al. *J. Phys. D: Appl. Phys.* 2008 **41** 135007
[70] A. Prikockytė, D. Bilc, P. Hermet, et al. *Phys. Rev. B* 2011 **84** 214301
[71] D. Crandles, F. Eftekhari, R. Faust, et al. *Appl. Optics* 2008 **47** 4205
[72] J. Hamlin, S. Deemyad, J. Schilling, et al. *Phys. Rev. B* 2007 **76** 014432
[73] J. Pietosa, B. Dabrowski, A. Wisniewski, et al. *Phys. Rev. B* 2008 **77** 104410
[74] X. Wan, J. Zhou, and J. Dong *Europhys. Lett.* 2012 **92** 57007
[75] S. Yamanaka, T. Maekawa, H. Muta, et al. *J. Solid State Chem.* 2004 **177** 3484
[76] N. Miao, B. Sa, J. Zhou, et al. *Comput. Mater. Sci.* 2011 **50** 1559
[77] T. Kiyama, K. Yoshimura, K. Kosuge, et al. *Phys. Rev. B* 1996 **54** 756
[78] C. Lee and X. Gonze *Phys. Rev. B* 1995 **51** 8610
[79] H. Hadipour and M. Akhavan *Eur. Phys. J. B* 2011 **84** 203
[80] B. J. Kennedy, C. J. Howard, and B. C. Chakoumakos *J. Phys.: Condens. Matter* 1999 **11** 1479

Table 1. The calculated lattice parameter $a$ (Å) and magnetic moment $\mu$ ($\mu_B$/f.u.) of cubic SrRuO$_3$ with different approaches in comparison with experimental data (Expt.).

| Code | Approach | $a$ (Å) | $\mu$ |
|---|---|---|---|
| ABINIT[a] | WC-GGA | 3.930 | 1.844 |
|  | LSDA | 3.888 | 1.399 |
| CRYSTAL[a] | B1-WC | 3.936 | 2.041 |
| VASP[b] | LSDA | 3.890 | 1.100 |
|  | LSDA+U | 3.890 | 1.640 |
| SIESTA[b] | LSDA | 3.890 | 1.260 |
|  | Pseudo-SIC | 3.890 | 1.770 |
| LMTO[c] | LSDA | 3.920 | 1.730 |
| LAPW[d] | LSDA | 3.923 | 1.170 |
| Expt.[e] | Fitted to 0 K | 3.905 | - |

[a]Present Work.

[b]Reference [25].

[c]Reference [18].

[d]Reference [16].

[e]Reference [8]. Experimental data linearly fitted to 0 K.

Table 2. Symmetry-adapted modes analysis (with the corresponding irreducible representations) of the different phases of SrRuO$_3$ after full atomic relaxation while keeping the cubic cell (no strain relaxation) within WC-GGA. Amplitudes in Å have been obtained with Amplimode [52,53]. ("-" indicates contributions forbidden by symmetry).

| Rotations | Symmetry | SPG NO. | Oxygen rotations | | Anti-polar motions | | Others oxygen motions | |
|---|---|---|---|---|---|---|---|---|
| | | | $R_4^+$ | $M_3^+$ | $X_5^+$ | $R_5^+$ | $M_2^+$ | $M_4^+$ |
| $a^0a^0c^-$ | $I4/mcm$ | 140 | 0.73 | - | - | - | - | - |
| $a^0a^0c^+$ | $P4/mbm$ | 127 | - | 0.71 | - | - | - | - |
| $a^0b^-b^-$ | $Imma$ | 74 | 0.80 | - | - | 0.08 | - | - |
| $a^0b^+b^+$ | $I4/mmm$ | 139 | - | 1.05 | - | - | - | 0.09 |
| $a^0b^+c^-$ | $Cmcm$ | 63 | 0.85 | 0.81 | 0.20 | 0.03 | - | 0.04 |
| $a^-b^+a^-$ | $Pnma$ | 62 | 1.00 | 0.77 | 0.35 | 0.06 | 0.00 | - |
| $a^+a^+c^-$ | $P4(2)/nmc$ | 137 | 1.15 | 1.18 | 0.30 | - | - | 0.07 |

Table 3. The calculated lattice parameters *a*, *b*, *c* in Å, Wyckoff positions and magnetic moments $\mu$ ($\mu_B$/f.u.) of distorted SrRuO$_3$ phases within WC-GGA in comparison with available experimental data. The rotation angle for the in-phase $M_3^+$ (+) tilt and the out-of-phase $R_4^+$ (-) tilt of the octahedra are indicated by $\phi^+$ and $\phi^-$ in degrees. The values in parentheses indicate the experimental data from Reference [12] for *Pnma*, Reference [8] for *I*4/*mcm*, and Reference [9] for *Imma*, respectively. The experimental data for the magnetic moments of *Pnma* are taken from Reference [60-62].

Table 3.

| | | | $a$ | $b$ | $c$ | Distortion Amplitude | $\phi^+/\phi^-$ | $\mu$ |
|---|---|---|---|---|---|---|---|---|
| $a^0a^0c^+$ | Cell(Å) | | 5.4990 | 5.4990 | 7.9715 | $\delta O_2$, 0.04992 | $c^+$: 11.26° | 1.87 |
| (*P4/mbm*) | Sr | 2(c) | 0 | 0.5 | 0.5 | | | |
| | Ru | 2(a) | 0 | 0 | 0 | | | |
| | $O_1$ | 2(b) | 0 | 0 | 0.5 | | | |
| | $O_2$ | 4(g) | 0.75+$\delta O_2$ | 0.25+$\delta O_2$ | 0.5 | | | |
| $a^0a^0c^-$ | Cell(Å) | | 5.4961 | 5.4961 | 7.9677 | $\delta O_2$, 0.05 (0.0255) | $c^-$: 11.53° | 1.87 |
| (*I4/mcm*) | | | (5.5784) | (5.5784) | (7.9078) | | ($c^-$: 5.85°) | |
| | Sr | 4(b) | 0 | 0.5 | 0.25 | | | |
| | Ru | 4(c) | 0 | 0 | 0 | | | |
| | $O_1$ | 4(a) | 0 | 0 | 0.25 | | | |
| | $O_2$ | 8(h) | 0.75+$\delta O_2$ | 0.25+$\delta O_2$ | 0 | | | |
| $a^0b^+c^-$ | Cell(Å) | | 7.7960 | 7.8654 | 7.8636 | $\delta Sr_{1y}$, 0.0124 | $c^-$: 8.93° | 1.94 |
| (*Cmcm*) | $Sr_1$ | 4(c) | 0 | $-\delta Sr_{1y}$ | 0.25 | $\delta Sr_{2y}$, 0.0136 | $b^+$: 8.41° | |
| | $Sr_2$ | 4(c) | 0 | 0.5$-\delta Sr_{2y}$ | 0.25 | $\delta O_{1x}$, 0.0385 | | |
| | Ru | 8(d) | 0.25 | 0.25 | 0 | $\delta O_{2y}$, 0.0360 | | |
| | $O_1$ | 8(e) | 0.25$-\delta O_{1x}$ | 0 | 0 | $\delta O_{2z}$, 0.0374 | | |
| | $O_2$ | 8(f) | 0 | 0.25+$\delta O_{2y}$ | 0.5+$\delta O_{2z}$ | $\delta O_{3x}$, 0.0398 | | |
| | $O_3$ | 8(g) | 0.75+$\delta O_{3x}$ | 0.25$-\delta O_{3y}$ | 0.25 | $\delta O_{3y}$, 0.0011 | | |
| $a^0b^-b^-$ | Cell(Å) | | 5.5052 | 7.7936 | 5.6186 | $\delta Sr_z$, 0.0053 | $a^-$: 8.49° | 1.93 |
| (*Imma*) | | | (5.564) | (7.874) | (5.597) | $\delta O_{1z}$, 0.0700 | | |
| | Sr | 4(e) | 0 | 0.25 | $-\delta Sr_z$ | $\delta O_{2y}$, 0.0391 | | |
| | Ru | 4(b) | 0 | 0 | 0.5 | | | |
| | $O_1$ | 4(e) | 0 | 0.25 | 0.5$-\delta O_{1z}$ | | | |
| | $O_2$ | 8(g) | 0.25 | $-\delta O_{2y}$ | 0.25 | | | |
| $a^-b^+a^-$ | Cell(Å) | | 5.5550 | 7.8370 | 5.5388 | $\delta Sr_x$, 0.0301 (0.0157) | $a^-$: 7.34° | 1.98 |
| (*Pnma*) | | | (5.5670) | (7.8446) | (5.5304) | $\delta Sr_z$, 0.0048 (0.0036) | ($a^-$: 6.23°) | (1.1-1.6) |
| | Sr | 4(c) | 0.5+$\delta Sr_x$ | 0.25 | 0.5+$\delta Sr_z$ | $\delta O_{1x}$, 0.0068 (0.0034) | $b^+$: 7.90° | |
| | Ru | 4(b) | 0 | 0 | 0.5 | $\delta O_{1z}$, 0.0630 (0.0532) | ($b^+$: 5.94°) | |
| | $O_1$ | 4(c) | 0.5$-\delta O_{1x}$ | 0.25 | $\delta O_{1z}$ | $\delta O_{2x}$, 0.0344 (0.0252) | | |
| | $O_2$ | 8(d) | 0.75$-\delta O_{2x}$ | 0.5+$\delta O_{2y}$ | 0.25+$\delta O_{2z}$ | $\delta O_{2y}$, 0.0326 (0.0278) | | |
| | | | | | | $\delta O_{2z}$, 0.0344 (0.0264) | | |

Table 4. The calculated Raman modes of orthorhombic SRO within WC-GGA in cm$^{-1}$ compared with available experimental data.

| Mode | WC-GGA | Expt.[a] | Expt.[b] | Expt.[c] | Expt.[d] | Expt.[e] |
|------|--------|----------|----------|----------|----------|----------|
| $A_g$ | 98 | - | - | - | 94 | 98 |
| $A_g$ | 131 | 123 | - | - | 118 | 130 |
| $A_g$ | 174 | 225 | - | - | - | - |
| $A_g$ | 225 | 252 | - | - | 221 | 230 |
| $A_g$ | 278 | 291 | - | - | 250 | 252 |
| $A_g$ | 383 | 393 | 373 | 372 | 388 | 393 |
| $A_g$ | 612 | - | - | - | - | - |
| $B_{1g}$ | 137 | - | - | - | - | - |
| $B_{1g}$ | 167 | - | - | - | - | - |
| $B_{1g}$ | 310 | - | - | - | - | - |
| $B_{1g}$ | 619 | - | - | - | - | - |
| $B_{1g}$ | 621 | - | - | - | - | - |
| $B_{2g}$ | 109 | - | - | - | - | - |
| $B_{2g}$ | 138 | - | - | - | - | - |
| $B_{2g}$ | 157 | - | - | - | - | - |
| $B_{2g}$ | 319 | - | - | - | - | - |
| $B_{2g}$ | 349 | 390 | 353 | 354 |  | 361 |
| $B_{2g}$ | 407 | 412 | 398 | 397 | 408 | 412 |
| $B_{2g}$ | 627 | - | - | - | - | - |
| $B_{3g}$ | 147 | - | - | - | - | - |
| $B_{3g}$ | 264 | - | - | - | - | - |
| $B_{3g}$ | 380 | - | - | - | - | - |
| $B_{3g}$ | 618 | - | - | - | - | - |
| $B_{3g}$ | 657 | - | - | - | - | - |

[a]Reference [65].

[b]Reference [67].

[c]Reference [68].

[d]Reference [64].

[e]Reference [66].

Table 5. The calculated IR modes of orthorhombic SRO within WC-GGA in cm$^{-1}$ compared with the available experimental data.

| Mode | WC-GGA | Expt.[a] | Expt.[b] | Expt.[c] |
|---|---|---|---|---|
| $B_{1u}$ | 112 | - | - | - |
| $B_{1u}$ | 159 | - | - | 150 |
| $B_{1u}$ | 236 | - | - | - |
| $B_{1u}$ | 243 | - | - | - |
| $B_{1u}$ | 283 | 285 | 285 | - |
| $B_{1u}$ | 318 | - | - | 300 |
| $B_{1u}$ | 348 | 380 | 328 | - |
| $B_{1u}$ | 364 | 435 | - | - |
| $B_{1u}$ | 555 | 580 | 580 | 552 |
| $B_{2u}$ | 141 | - | - | - |
| $B_{2u}$ | 153 | - | - | - |
| $B_{2u}$ | 239 | - | - | - |
| $B_{2u}$ | 277 | - | - | - |
| $B_{2u}$ | 339 | - | - | - |
| $B_{2u}$ | 532 | - | - | - |
| $B_{2u}$ | 555 | - | - | - |
| $B_{3u}$ | 118 | - | - | - |
| $B_{3u}$ | 155 | - | - | - |
| $B_{3u}$ | 217 | - | - | - |
| $B_{3u}$ | 232 | - | - | - |
| $B_{3u}$ | 270 | - | - | - |
| $B_{3u}$ | 276 | 285 | 285 | - |
| $B_{3u}$ | 382 | 328 | 380 | - |
| $B_{3u}$ | 424 | - | 435 | - |
| $B_{3u}$ | 541 | 580 | 580 | - |

[a]Reference [69].

[b]Reassignment of experimental data from reference [69].

[c]Reference [71].

Table 6. The calculated rotation angles in degrees relative to the ideal cubic $SrRuO_3$ for different phases with: (1) oxygen relaxation only (cation and cell fixed to cubic), $\phi_1$; (2) oxygen and cation relaxation (full atomic relaxation, keeping the unit cell fixed to cubic), $\phi_2$; (3) full atomic and strain relaxations, $\phi_3$.

| Rotations | Symmetry | SPG NO. | $\phi^+_1$ | $\phi^-_1$ | $\phi^+_2$ | $\phi^-_2$ | $\phi^+_3$ | $\phi^-_3$ |
|---|---|---|---|---|---|---|---|---|
| $a^0a^0a^0$ | $Pm\bar{3}m$ | 221 | 0.00 | 0.00 | 0.00 | 0.00 | 0.00 | 0.00 |
| $a^0a^0c^-$ | $I4/mcm$ | 140 | 0.00 | -10.68 | 0.00 | -11.52 | 0.00 | -11.53 |
| $a^0a^0c^+$ | $P4/mbm$ | 127 | 10.28 | 0.00 | 10.28 | 0.00 | 11.26 | 0.00 |
| $a^0b^-b^-$ | $Imma$ | 74 | 0.00 | 8.12 | 0.00 | 8.29 | 0.00 | 8.49 |
| $a^0b^+b^+$ | $I4/mmm$ | 139 | 7.68 | 0.00 | 7.67 | 0.00 | 7.75 | 0.00 |
| $a^0b^+c^-$ | $Cmcm$ | 63 | 7.33 | 8.68 | 8.37 | 8.76 | 8.41 | 8.93 |
| $a^-b^+a^-$ | $Pnma$ | 62 | 6.56 | 6.53 | 7.89 | 7.30 | 7.90 | 7.34 |
| $a^-a^-a^-$ | $R\bar{3}c$ | 167 | 0.00 | 6.54 | 0.00 | 6.54 | 0.00 | 6.54 |
| $a^+a^+c^-$ | $P4(2)/nmc$ | 137 | 5.14 | 8.50 | 6.07 | 8.36 | 5.84 | 8.80 |

Table 7. The calculated gains of energy in meV/f.u., respect to the ideal cubic SrRuO$_3$ phase taken as reference, for different relaxed phases, labelled in terms of the compatible tilt pattern. Please refer to the caption of Figure 4 for the definitions of different energies.

| Rotations | Symmetry | SPG NO. | $E'_{oxygen}$ | $E_{oxygen}$ | $E_{cation}$ | $E_{strain}$ |
|---|---|---|---|---|---|---|
| $a^0a^0a^0$ | $Pm\bar{3}m$ | 221 | 0 | 0 | 0 | 0 |
| $a^0a^0c^-$ | $I4/mcm$ | 140 | 151 | 151 | 0 | 16 |
| $a^0a^0c^+$ | $P4/mbm$ | 127 | 131 | 131 | 0 | 16 |
| $a^0b^-b^-$ | $Imma$ | 74 | 173 | 181 | 7 | 9 |
| $a^0b^+b^+$ | $I4/mmm$ | 139 | 143 | 153 | 0 | 2 |
| $a^0b^+c^-$ | $Cmcm$ | 63 | 160 | 173 | 20 | 5 |
| $a^-b^+a^-$ | $Pnma$ | 62 | 159 | 181 | 37 | 2 |
| $a^-a^-a^-$ | $R\bar{3}c$ | 167 | 182 | 182 | 0 | 16 |
| $a^+a^+c^-$ | $P4(2)/nmc$ | 137 | 163 | 170 | 22 | 2 |

Table 8. Selective lowest-order invariant polynomials [54] of the Landau-type energy expansion around the cubic phase of SrRuO$_3$ involving the modes shown in Figure 4. Only the lowest-order anharmonic terms involving the coupling with a mode at the linear order and responsible for improper-type behavior are reported.

| Rotations | Symmetry | SPG NO. | Invariant Polynomials for Coupling Modes |
|---|---|---|---|
| $a^-b^0a^-$ | *Imma* | 74 | $Q(R_4^+)Q(R_5^+)^3+Q(R_4^+)^3Q(R_5^+)...$ |
| $a^0c^+c^+$ | *I4/mmm* | 139 | $Q(M_4^+)Q(M_3^+)^2...$ |
| $a^0b^+c^-$ | *Cmcm* | 63 | $Q(R_4^+)Q(M_3^+)Q(X_5^+)+Q(R_5^+)Q(M_3^+)Q(X_5^+)+Q(R_4^+)Q(M_4^+)Q(X_5^+)...$ |
| $a^-b^+a^-$ | *Pnma* | 62 | $Q(R_4^+)Q(M_3^+)Q(X_5^+)+Q(R_4^+)Q(M_2^+)Q(X_5^+)+Q(R_5^+)Q(M_3^+)Q(X_5^+)...$ |
| $a^+a^+c^-$ | *P*4(2)/*nmc* | 137 | $Q(X_5^+)^2Q(M_4^+)+Q(R_4^+)Q(M_4^+)Q(X_5^+)+Q(R_4^+)Q(M_3^+)Q(X_5^+)...$ |

Table 9. The calculated lattice parameters *a*, *b*, *c* in Å, reduced atomic position as well as the magnetic moment $\mu$ ($\mu_B$/f.u.) of *Pnma* SrRuO$_3$ phases in comparison with available experimental and other theoretical data.

|  | ABINIT WC-GGA | ABINIT LSDA | Expt.[a] | CRYSTAL[b] B1-WC | SIESTA[b] LSDA+U | VASP[c] LSDA | VASP[d] LSDA |
|---|---|---|---|---|---|---|---|
| *a* | 5.5550 | 5.4997 | 5.5670 | 5.5696 | 5.5247 | 5.4924 | 5.5031 |
| *b* | 5.5388 | 5.4874 | 5.5304 | 5.5279 | 5.4881 | 5.4887 | 5.4828 |
| *c* | 7.8370 | 7.7542 | 7.8446 | 7.8392 | 7.7759 | 7.7561 | 7.7546 |
| Sr *x* | -0.0048 | -0.0053 | -0.0027 | -0.0036 | -0.0041 | -0.0050 | -0.0050 |
| Sr *y* | 0.0301 | 0.0303 | 0.0157 | 0.0276 | 0.0274 | 0.0304 | 0.0296 |
| Sr *z* | 0.2500 | 0.2500 | 0.2500 | 0.2500 | 0.2500 | 0.2500 | 0.2500 |
| Ru *x* | 0.5000 | 0.5000 | 0.5000 | 0.5000 | 0.5000 | 0.5000 | 0.5000 |
| Ru *y* | 0.0000 | 0.0000 | 0.0000 | 0.0000 | 0.0000 | 0.0000 | 0.0000 |
| Ru *z* | 0.0000 | 0.0000 | 0.0000 | 0.0000 | 0.0000 | 0.0000 | 0.0000 |
| O(1) *x* | 0.0630 | 0.0642 | 0.0532 | 0.0612 | 0.0639 | 0.0650 | 0.0647 |
| O(1) *y* | 0.4932 | 0.4924 | 0.4966 | 0.4943 | 0.4953 | 0.4942 | 0.4941 |
| O(1) *z* | 0.2500 | 0.2500 | 0.2500 | 0.2500 | 0.2500 | 0.2500 | 0.2500 |
| O(2) *x* | 0.7156 | 0.7163 | 0.7248 | 0.7170 | 0.7837 | 0.7158 | 0.7165 |
| O(2) *y* | 0.2843 | 0.2835 | 0.2764 | 0.2837 | 0.2162 | 0.2834 | 0.2834 |
| O(2) *z* | 0.0326 | 0.0334 | 0.0278 | 0.0313 | 0.0330 | 0.0340 | 0.0336 |
| $\mu$ | 1.98 | 1.70 | 1.1-1.6 | 2.00 | 2.00 | 0.79 | 0.98 |

[a]Reference [12] and magnetic moments from reference [60-62].

[b]Reference [6].

[c]Reference [25].

[d]Reference [20].

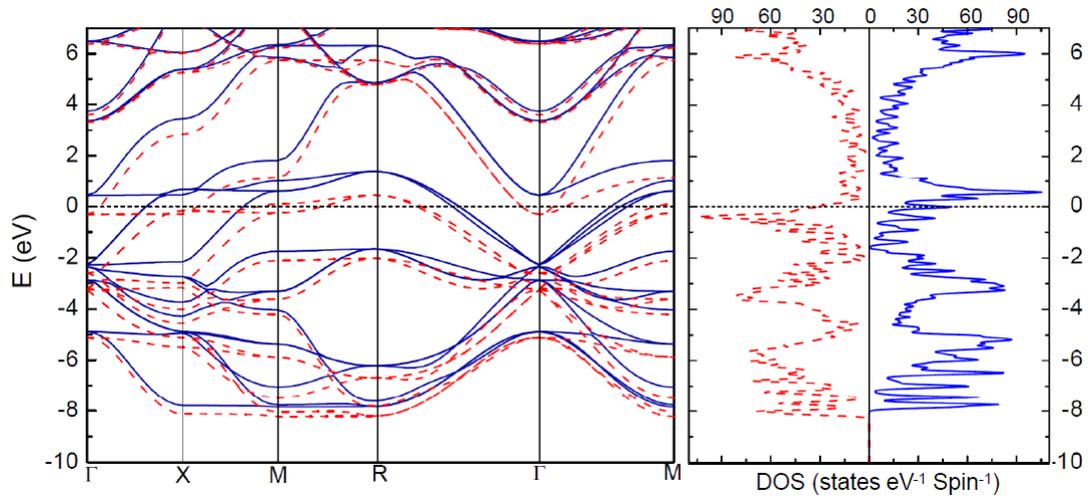

Figure 1. The calculated electronic band structure and density of states within WC-GGA for the relaxed cubic SrRuO$_3$. The red dash lines denote the majority spin, while the blue solid lines refer to the minority spin. The Fermi level is set to be zero.

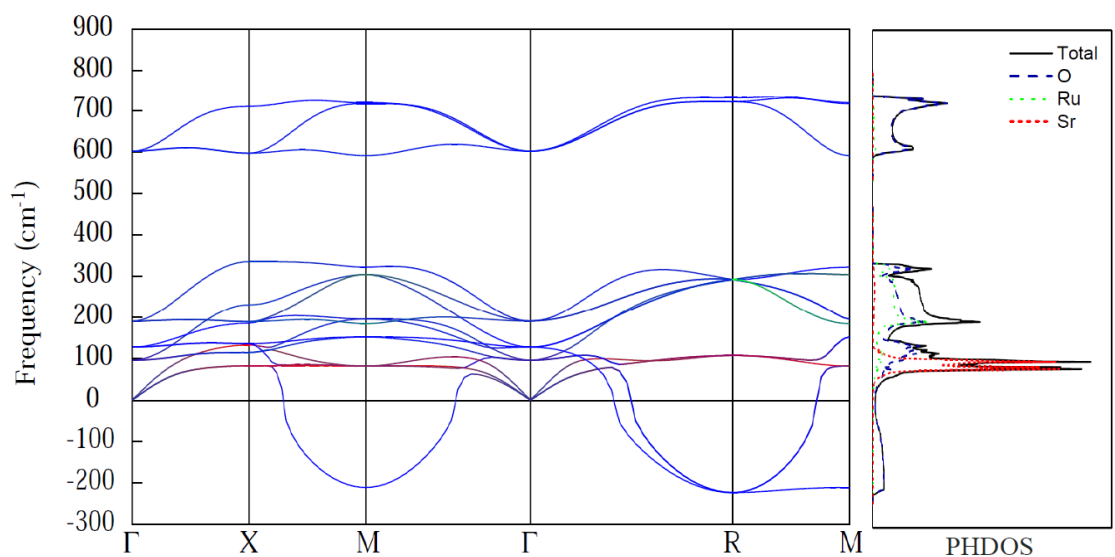

Figure 2. The calculated phonon dispersion curves and phonon density of states of cubic SrRuO$_3$ within WC-GGA along the *Γ-X-M-Γ-R-M* of the cubic Brillouin zone. The total DOS and the projected DOS of oxygen, ruthenium and strontium atoms are plotted using solid line (in black), dash line (in blue), dot line (in green) and short dash line (in red), respectively. Negative frequencies indicate imaginary values.

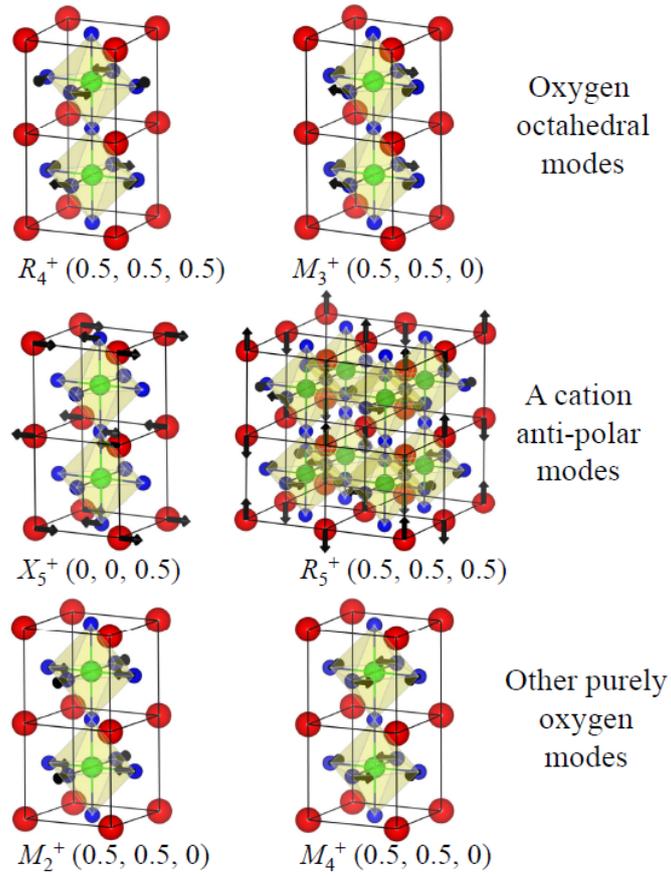

Figure 3. A schematic illustration of relevant phonon modes of SrRuO$_3$. Black arrows indicate the atomic motions. Sr atoms (in red) at the corners, Ru atoms (in green) at the centers and O atoms (in blue) at the face centers of the perovskite cubic cell. The wave vectors are also given in the parentheses for corresponding modes.

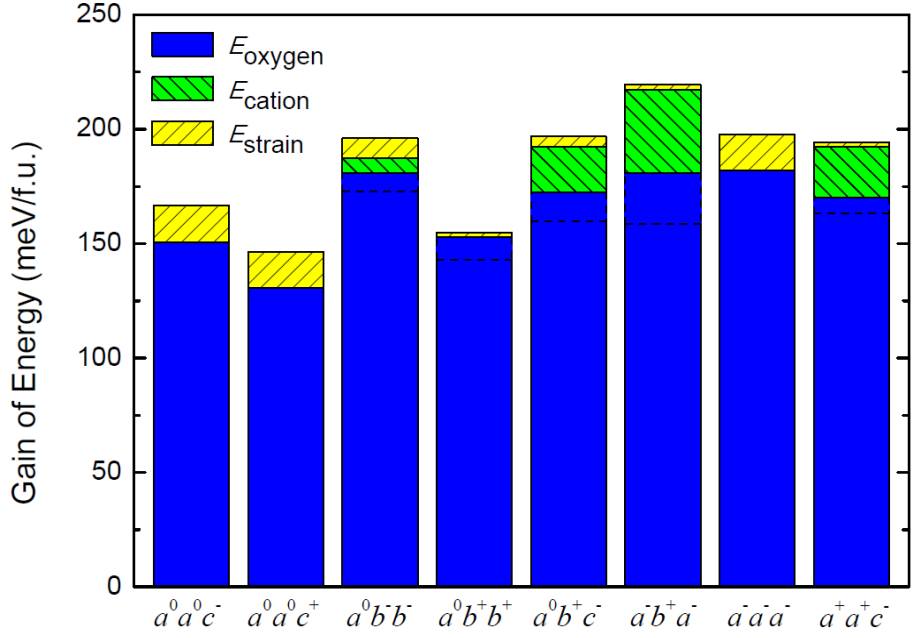

Figure 4. The calculated gains of energy, respect to the ideal cubic SrRuO$_3$ phase taken as reference, for different relaxed phases, labelled in terms of the compatible tilt pattern. $E_{oxygen}$ corresponds to the gain of energy that can be achieved from the relaxation of oxygen atomic positions only. $E_{cation}$ corresponds to the supplemental gain of energy that can be achieved when allowing for additional concomitant cation motions. In this latter case, the oxygen distortions are modified through the coupling with cation motions : the dashed line identifies the reduced gain of energy ($E'_{oxygen}$) produced by pure oxygen motions in this fully relaxed phase. All the previous calculations are done when keeping the unit cell fixed. $E_{strain}$ corresponds to the additional gain of energy when allowing for simultaneous strain relaxation. The sum ($E_{oxygen}+E_{cation}+E_{strain}$) is the maximum gain of energy than can be achieved from full structural relaxation for each phase.

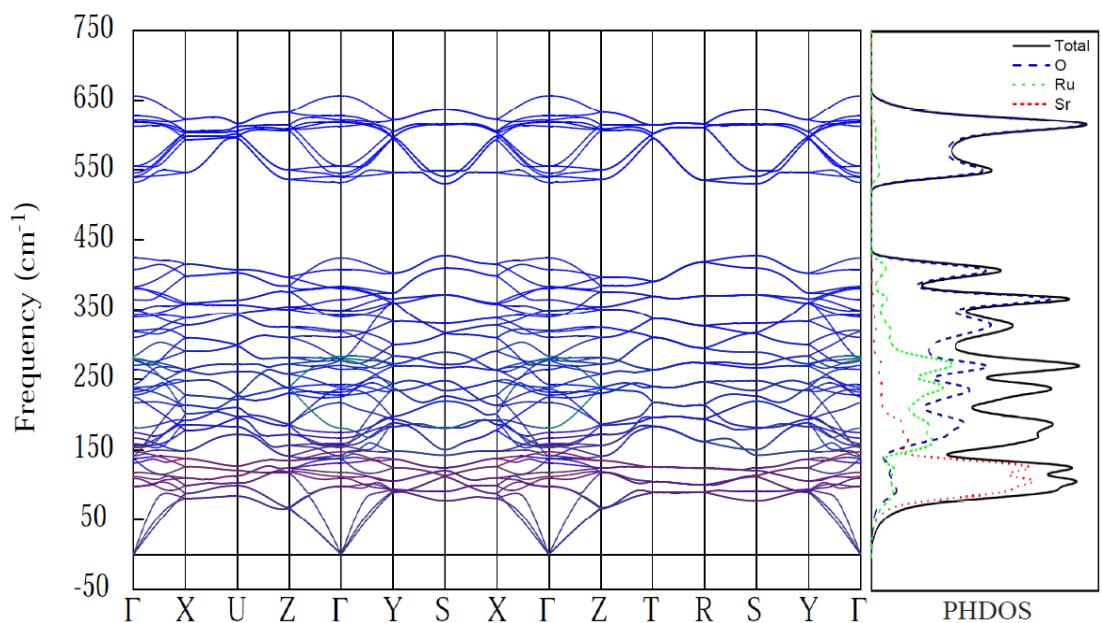

Figure 5. The calculated full phonon dispersion curves and phonon density of states of ground state SrRuO$_3$ within WC-GGA. The solid line (black), dash line (blue), dot line (green) and short dash line (red) represent the total DOS and the projected DOS of oxygen, ruthenium and strontium atoms, respectively. Negative frequencies indicate imaginary values.

Figure 6. The calculated heat capacity $C_V$ for the *Pnma* phase of SrRuO$_3$ (4 f.u., 20 atoms) within WC-GGA compared with the available experimental data measured at 15-250K and 300-680K from Reference [17] and Reference [75], respectively.

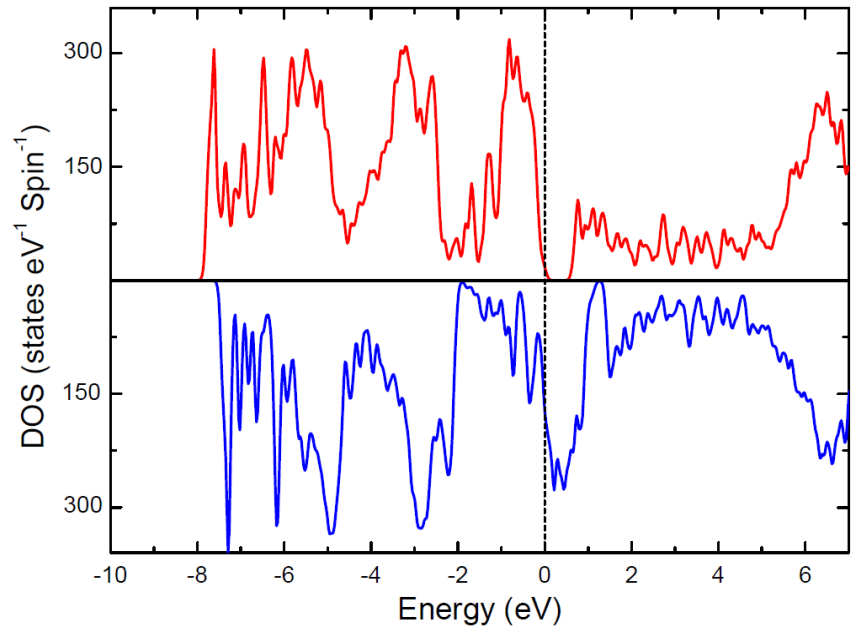

Figure 7. The calculated electronic density of states for majority spins (up, red) and minority spins (down, blue) within WC-GGA for orthorhombic $SrRuO_3$. The Fermi level is set to be zero.

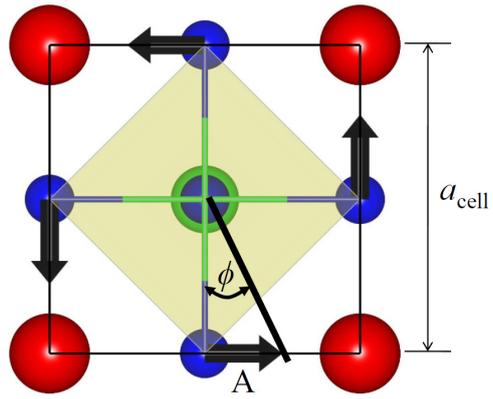

Figure 8. The amplitude of the rotation angle ($\phi$) is deduced from the amplitude A of the overlap of the distorted structure with $R_4^+$ and $M_3^+$ modes according to $\phi=\arctan(2A/a_{cell})$. Sr atoms (in red) at the corners, Ru atoms (in green) at the centers and O atoms (in blue) at the face centers of the perovskite cubic cell.